\documentclass[preprint,showpacs,preprintnumbers,amsmath,amssymb]{revtex4}

\usepackage{graphicx}
\usepackage{dcolumn}
\usepackage{bm}

\begin{document}

\preprint{}

\title{Discernment of Hubs and Clusters in Socioeconomic Networks}

\author{Paul B. Slater}%
\email{slater@kitp.ucsb.edu}
\affiliation{%
ISBER, University of California, Santa Barbara, CA 93106\\
}%
\date{\today}

\begin{abstract}
Interest in the analysis of networks has grown rapidly 
in the new millennium. Consequently, we
promote renewed attention to a certain methodological approach  introduced
in 1974. Over the succeeding decade, 
this {\it two}-stage--{\it double}-standardization 
and hierarchical clustering 
(single-linkage-like)--procedure was applied to a wide variety of
weighted, directed networks of a socioeconomic nature, 
frequently revealing the presence of ``hubs''. 
These were, typically--in the numerous instances studied of 
migration flows between 
geographic subdivisions within
nations--``cosmopolitan/non-provincial'' areas, a prototypical example being 
the French capital, 
Paris. Such locations emit and absorb people broadly across their 
respective nations. Additionally, the two-stage 
procedure--which ``might very well be the most successful application
of cluster analysis'' (R. C. Dubes, 1985)--detected many
(physically or socially) isolated, functional groups 
(regions)
of areas, such as the southern
islands, Shikoku and Kyushu, of Japan, the Italian islands 
of Sardinia and Sicily, 
and the New England region of the 
United States. Further, we discuss a (complementary)
approach developed in 1976, in which the {\it max-flow/min-cut} theorem
was applied
to {\it raw/non-standardized} (interindustry, as well as migration) flows.
\end{abstract}


\maketitle
\section{Introduction}
A. L. Barab{\' a}si, in his 
recent popular book, ``Linked'', asserts that the emergence 
of {\it hubs} in networks is a surprising phenomenon that is ``forbidden by both the Erd{\"o}s-R{\'e}nyi and Watts-Strogatz models" \cite[p. 63]{linked} 
\cite[Chap. 8]{siegfried}.
Here, we indicate an analytical framework introduced
in 1974 that the distinguished computer scientist 
R. C. Dubes, in a review of the 
compilation of multitudinous results  \cite{tree}, asserted 
``might very well be the most successful application 
of cluster analysis'' \cite[p. 142]{dubes}. 
This {\it two}-stage methodology has proved insightful
in revealing--among other interesting relationships--hub-like structures 
in networks of (weighted, directed) internodal flows. 
This approach, together with its many diverse socioeconomic
applications, was documented in a large number of 
(subject-matter and technical) journal articles
(among them \cite{japan,france,winchester,science,partial,IO,college,gentileschi,metron,SEAS,qq,spain,schwarz,romania,russia,hirst1,seps,india}), 
as well as in the research institute monographs 
\cite{tree,county,tree2}. It has also been the subject of various
comments, criticisms  and discussions 
\cite{masser,hirst2,findlay,seps,holmes1,holmes2,holmes3,telephone,baumann,clark,boyd,pandit,hoover} (cf. \cite{noronha,corvers}).

Though the principal procedure to be detailed here is applicable 
in a wide variety of social-science 
settings \cite{tree,dubes}, it has been primarily 
used, in a demographic context, to study the 
{\it internal migration} tables published at regular periodic intervals by most of the nations of the world. 
These tables can be thought of as $N \times N$ (square) matrices, the entries ($m_{ij}$) of which are the number of people who
lived in geographic subdivision $i$ at time $t$ and $j$ at 
time $t+1$. (Some tables--but not all--have diagonal entries, $m_{ii}$, which may represent either the number of people who did move within
area $i$, or simply those who lived in $i$ both 
at $t$ and $t+1$. It can 
sometimes be of interest to compare analyses with zero  and
nonzero diagonal entries \cite{county}.)
\section{Two-Stage Methodology}
\subsection{First Step: Double-Standardization} \label{first}
In the {\it first} step (iterative proportional fitting procedure [IPFP] 
\cite{fienberg}) 
of the methodology under discussion here, 
the rows and columns of the 
table of flows 
are alternately  (biproportionally \cite{bacharach}) scaled 
to sum to a fixed number (say 1). Under broad conditions--to be discussed 
below--convergence occurs to a ``doubly-stochastic" 
(bistochastic) 
table, with row and column sums all 
simultaneously equal to 1 \cite{mosteller,louck,CSBZ,unistochastic,
romney,wong}. 
The purpose of the scaling is to 
remove overall (marginal) effects of size, and focus on relative, 
interaction  effects. 
Nevertheless, the {\it cross-product ratios} 
({\it relative odds}), $\frac{m_{ij} m_{kl}}{m_{il} m_{kj}}$, 
measures of association, are left {\it invariant}. 
Additionally, the entries of the
doubly-stochastic table provide 
{\it maximum  entropy} estimates of the original
flows, given the row and column constraints \cite{eriksson,macgill}.

For large {\it sparse} 
flow tables, only the nonzero entries, together with their
row and column coordinates are needed. Row and column (biproportional) 
multipliers can be iteratively computed by sequentially accessing the nonzero
cells \cite{parlett}. If the table is ``critically sparse'', various convergence difficulties may occur. Nonzero entries that are ``unsupported''--that is, not part of a set of $N$ nonzero entries, no two in the same row and 
column-- may converge to zero and/or the 
biproportional multipliers may not converge \cite[p. 19]{tree} \cite{sinkhorn} \cite[p. 171]{mirsky}.
The ``first strongly polynomial-time algorithm for matrix scaling'' 
was reported in \cite{linial}.

The scaling was successfully implemented, in our largest analysis,  
with a
$3,140 \times 3,140$ 1965-70 intercounty migration table--having 94.5\% 
of its entries, zero--for
the United States \cite{county,partial}, as well as for a more aggregate
$510 \times 510$ table (with {\it State Economic Areas} as  the basic 
unit) for the US for the same period \cite{SEAS}. ({\it Smoothing} procedures
could be used to modify the zero-nonzero structure 
of a flow table, particularly 
if it is critically sparse \cite{simonoff,boundary}. If one takes the 
second power of 
a doubly-stochastic matrix, one obtains another such 
matrix, but smoother in character. One might also 
consider standardizing the {\it i}th row [column] sum 
to be proportional to the number of non-zero entries in the 
{\it i}th row [column].)
\subsection{Second Step: Strong Component Hierarchical Clustering}
In the {\it second} step of the two-stage 
procedure, the doubly-stochastic matrix 
is converted to a series of {\it directed} 
(0,1) graphs (digraphs), by applying thresholds to its entries. 
As the thresholds   are 
progressively lowered, larger and larger {\it strong components} 
(a directed path existing from any member of a component 
to any other) of the resulting 
graphs are found. This process 
(a simple variant of well-known single-linkage [nearest-neighbor or min]
clustering \cite{gower1}) can be represented by the familiar dendrogram 
or tree diagram used in 
hierarchical cluster analysis and cladistics/phylogeny (cf. \cite{ozawa,hubert}).
\subsection{Computer implementations}
A FORTRAN implementation of the two-stage process was given in 
\cite{leusmann}, as well as a realization 
in the SAS (Statistical Analysis System) 
framework \cite{chilko}. Subsequently, 
the noted computer scientist R. E. Tarjan 
\cite{schwartz} devised an $O(M (\log{N})^2)$
algorithm \cite{tarjan} and, then, a further improved $O(M (\log{N}))$ 
method \cite{tarjan2}, 
where $N$ is the number of nodes and $M$ the number of edges of 
a directed graph. (These substantially improved upon 
the earlier works \cite{leusmann,chilko}, 
which 
required the 
computations of {\it transitive closures} of graphs, and were 
$O(M N)$ in nature.) A FORTRAN coding--involving 
linked lists--of the improved Tarjan 
algorithm \cite{tarjan2} was presented
in \cite{tarjanslater}, and applied in the 
aforementioned US intercounty study \cite{county}. 
If the graph-theoretic (0,1)-structure of a network under study 
is {\it not} strongly connected 
\cite{hartfiel}, {\it independent} two-stage analyses of 
the subsystems of the network would be appropriate.
\subsection{Goodness-of-fit}
The {\it goodness-of-fit} of the dendrogram generated 
to the doubly-stochastic table itself
can be evaluated--and possibly employed, it would seem, as an optimization 
criterion (cf. \cite[p. 210]{hansen} \cite[sec. 3]{cmn}). In the 
context--not 
of the weighted, directed networks under discussion here--but 
of (0,1)-networks or simply graphs,
Clauset, Moore and Newman have written: 
``[t]he method known as \emph{hierarchical clustering} groups vertices in networks by aggregating them iteratively in a hierarchical fashion.  However, it is not clear that the hierarchical structures produced by these and other popular methods are unbiased, as is also the case for the hierarchical clustering algorithms of machine learning.  That is, it is not clear to what degree these structures reflect the true structure of the network, and to what degree they are artifacts of the algorithm itself. This conflation of intrinsic network properties with features of the algorithms used to infer them is unfortunate \ldots 
we give a precise definition of hierarchical structure, give a generic model for generating arbitrary hierarchical structure in a random graph, and describe a statistically principled way to learn the set of hierarchical features that most plausibly explain a particular real-world network''. \cite{cmn}.

Distances between nodes 
in the dendrogram satisfy the (stronger than {\it triangular}) 
{\it ultrametric} 
inequality, $d_{ij} \leq \max{(d_{ik},d_{jk})}$ \cite[p. 245]{johnson} 
\cite[eq. (2.2)]{rammal}.
\section{Empirical Results}
\subsection{Cosmopolitan or Hub-Like Units}
\subsubsection{Internal migration flows}
Geographic subdivisions (or groups of subdivisions) that enter into the 
bulk of the dendrogram at the weakest levels are those with the broadest ties. Typically, these have been found to be 
``cosmopolitan", hub-like areas, 
a prototypical example being 
the French capital, Paris \cite[sec. 4.1]{tree} \cite{france}. 
Similarly, 
in parallel analyses of other 
internal migration tables, the cosmopolitan/non-provincial natures
of London \cite{siegen}, 
Barcelona \cite{spain} \cite[sec. 6.2, Figs. 36, 37]{tree}, 
Milan \cite{gentileschi} \cite[sec. 6.3, Figs. 39, 40]{tree} 
(cf. \cite{metron}), Amsterdam 
\cite[p. 78]{tree} \cite{masser},
West Berlin \cite[p. 80]{tree}, Moscow (the city and the oblast as a unit) 
\cite{russia} 
\cite[sec. 5.1 and Figs. 6, 7]{tree}, Manila (coupled with suburban Rizal) 
\cite{manila}, 
Bucharest \cite{romania}, 
{\^I}le-de-Montr{\'e}al \cite[p. 87]{tree}, 
Z{\"u}rich, Santiago, Tunis and Istanbul \cite{turkish} 
were--among
others--highlighted in the respective dendrograms for their nations
 \cite[sec. 8.2]{tree} 
\cite[pp. 181-182]{qq} \cite[p. 55]{science}. In the intercounty analysis
for the US, the most cosmopolitan entities were: (1) the 
{\it centrally} located paired
Illinois counties of Cook (Chicago) and neighboring, suburban Du Page; 
(2) the nation's capital, Washington, D. C.; and (3) the paired south
Florida (retirement) counties of Dade (Miami) and Broward (Ft. Lauderdale) 
\cite{county,partial,fields}. In general, counties with large military 
installations, large college populations or state capitals 
also interacted broadly with other areas \cite[p. 153]{county}. 
Application of the two-stage methodology to 1965-66 London inter-borough
migration \cite{masser} indicated that the three inner boroughs of Kensington
and Chelsea, Westminster, and Hammersmith acted--as a unit--in a 
cosmopolitan manner \cite[sec. 5.2, Fig. 10]{tree}. 
(In sec. 8.2 and Table 16 of the anthology of results \cite{tree}, 
additional geographic units and groups of
units found to be cosmopolitan with regard to migration, are enumerated.)

It should be emphasized that 
although the indicated cosmopolitan areas may generally have 
relatively large populations, 
this can not, in and of itself, 
explain the wide national ties observed, since the 
double-standardization, in effect, renders all areas of equal overall size.
(However, to the extent that larger areas do have fewer zero entries in their
corresponding rows and columns, a bias to cosmpolitanism may 
in fact be present, 
which should be carefully considered. Possible corrections for bias were 
discussed above in sec.~\ref{first}.)
If one were to obtain a (zero-diagonal) 
doubly-stochastic matrix, all the entries of which
were simply $\frac{1}{N-1}$, it would indicate complete 
indifference among migrants
as to where they come from and to where they go.
A maximally cosmopolitan unit would be one for which all the corresponding
row and column entries were $\frac{1}{N-1}$ (if all the diagonal
entries, $m_{ii}$, are {\it a priori} zero).
(It seems interesting to note that cosmpolitan areas appear to have
a certain {\it minimax} character, that is, the maximum doubly-stochastic
entry for the corresponding row and column tends to be minimized.)
\subsubsection{Trade and interindustry flows}
The nation of Italy possessed the broadest ties in a two-stage analysis
of the value of 1974 trade between 113 nations, followed by a closely-bound
group composed of the four Scandinavian countries \cite{schwarz} 
\cite[sec. 5.6, Fig. 22]{tree}.
In a two-stage 
study (but using {\it weak} rather than strong components of the 
associated digraphs) of the 
1967 US interindustry transaction table, the industry
with the broadest (most diffuse) ties was found to be Other Fabricated
Metal Products \cite{IO,gosling} \cite[pp. 13-18]{tree2}.
\subsubsection{Journal citations}
 In a two-stage analysis of 
22 mathematical journals, the {\it Annals of Mathematics} and {\it 
Inventiones Mathematicae} were strongly paired, while the {\it Proceedings of
the American Mathematical Society} was found to 
possess the broadest, most diffuse ties 
\cite{science}.

In a recent, large-scale ($N>6000$) journal-to-journal 
 citation analysis, decomposing ``the network into modules by compressing 
a description of the probability flow'', Rosvall and Bergstrom 
preliminarily {\it omitted } from their analysis 
the prominent journals {\it Science}, 
{\it Nature} and {\it Proceedings of the National Academy of Sciences} 
 \cite[p. 1123]{rosvall}.
(Those are precisely the ones that would be expected to be ``cosmopolitan'' 
or hub-like in 
character, and to be highlighted in a corresponding
two-stage analysis.) Their rationale for 
the omission was that ``the broad scope of these journals otherwise creates an 
illusion of tighter connections among disciplines, when in fact few readers 
of the physics articles in {\it Science} also are close readers of the 
biomedical articles therein''. (In \cite[pp. 125-153]{tree2}, we reported
the results of a {\it partial} hierarchical clustering--not 
a two-stage analysis, but one originally designed and conducted 
by Henry G. Small and William 
Shaw--of citations between more than 3,000 journals. The clusters 
obtained there were compared with the actual 
subject matter classification employed
by the Institute for Scientific Information.)
\subsection{Functional Clusters of Units}
\subsubsection{Internal migration regions}
Geographically isolated (insular) areas--such as the Japanese islands of 
Kyushu and Shikoku \cite{japan}--emerged 
as well-defined {\it clusters} (regions)
of their constituent (seven and four, respectively) 
subdivisions (``prefectures'' in the Japanese case) 
in the dendrograms for the two-stage analyses, and similarly
 the Italian islands of Sicily and Sardinia 
\cite{gentileschi}, the North and South Islands of New Zealand, and the
Canadian islands of 
Newfoundland and Prince Edward Island \cite[p. 90]{tree}  
(cf. \cite{e,multiterminal}).
The eight counties of Connecticut, and other New England groupings, as  
further examples,  were
also very prominent in the highly disaggregated US analysis \cite{county}. 
Relatedly, in a study based solely upon 
the 1968 movement of {\it college students} among
the fifty states, the six New England states were strongly clustered 
\cite[Fig. 1]{college}. Employing a 1963 Spanish interprovincial migration
table, well-defined regions were formed by the two provinces of
the Canary Islands, and the four provinces of Galicia \cite{spain} 
\cite[sec. 6.2.1, Fig. 37]{tree}. 
The southernmost Indian states of Kerala and Madras (now Tamil Nadu) 
were strongly paired on the basis of 1961 interstate flows \cite{india}.
A detailed comparison between functional migration regions found by 
the two-stage procedure and those actually 
employed for administrative, political 
purposes in the corresponding nations is given in sec. 8.1 and Table 15 of 
\cite{tree}. 

It should be noted that it is rare  that the two-stage
methodology yields a migration region 
composed of two or more noncontiguous subregions--even though no contiguity
information at all is present in the flow table 
nor provided to the algorithm (cf. \cite{loglinear,boundary}).
A notable exception to this rule was the uniting of the northern 
Italian region of
Piemonte--the location of industrial Turin, where Fiat is based--with 
southern regions, {\it before} joining with central regions, in an 
18-region  1955-70 study \cite{metron} 
\cite[p. 75]{tree} (cf. \cite{gentileschi}).
\subsubsection{Intermarriage and interindustry clusters}
 In a two-stage analysis of a $32 \times 32$ table of 
birthplace of bridegroom versus birthplace of bride of 1947 Australian
intermarriages \cite{price}, Greece and Cyprus were the strongest dyad 
\cite[sec. 5.7, Fig. 25]{tree}.

In the 1967 US interindustry 
two-stage ({\it weak} component) analysis, two particularly salient pairs 
of functionally-linked industries were: (1)
Stone and Clay Products, and Stone and Clay Mining and Quarrying; and (2) 
Household Appliances and Service Industry Machines (the latter industry
purchases laundry equipment, refrigerators and freezers from the former)
\cite{IO,gosling} \cite[pp. 13-18]{tree2}. 

\section{Statistical Aspects}
It would be of interest to develop a theory--making use of 
the rich mathematical structure of doubly-stochastic 
matrices--by which the {\it statistical
significance} of apparent hubs and clusters 
in dendrograms produced by the two-stage procedure 
could be evaluated \cite[pp. 7-8]{county} \cite{bock}. 
In the geographic context of internal migration tables, where nearby areas 
have a strong distance-adversion predilection for binding, it seems unlikely 
that most clustering
results generated could be considered to be--in any standard 
sense--``random'' in nature. 
On the other hand, other types of ``origin-destination'' 
tables, such as those for 
{\it occupational} mobility \cite{duncan}, journal citations 
\cite{science} \cite[pp. 125-153]{tree2}, interindustry (input-output) flows 
\cite{IO} \cite{gosling}, brand-switches \cite[sec. 9.6]{tree} \cite{rao}, 
crime-switches \cite[sec. 9.7]{tree} \cite[Table XII]{blumstein}, and (Morse code) 
confusions \cite[sec. 9.8]{tree} \cite{rothkopf}, among others, clearly lack such a geographic dimension (cf. \cite{point}). 
An efficient algorithm--considered as a nonlinear dynamical system--to generate {\it random} bistochastic matrices has
recently been presented \cite{CSBZ} (cf. \cite{griffiths,ZKSS}).

In the US 3,140-county migration study, a statistical 
test of Ling \cite{ling}  (designed 
for {\it undirected} graphs), based on the difference in
the ranks of two edges, was employed in a heuristic manner
\cite[pp. 7-8]{county}. 
For example, the 3,148th largest doubly-stochastic value, 0.12972 
(corresponding to the flow from Maui County to Hawaii County), {\it united}
the four counties of the state of Hawaii. The (considerably weaker) 
7,939th largest value, 
0.07340 (the link from Kauai County, Hawaii, to Nome, Alaska), {\it integrated}
the four-county 
state of Hawaii into a much larger 
2,464-county cluster. The difference of these
two ranks, 4,192 =  7,340 - 3,148, is the isolation index or ``survival
time'' of this state as a cluster. Reference to Table 1 in \cite{county} 
showed the significance of the state of 
Hawaii as a functional 
internal migration unit at the 0.01 level \cite[p. 7]{county}. 
(In the computation of this table, the approximation was used that
the number of edges in the relevant 
digraphs was a negligible proportion of all 
possible $3,140 \times 3,139$ edges.)

Also, the possibility of
employing the {\it asymptotic} theory of random digraphs 
\cite{palasti,karonski} for statistical testing purposes was raised 
in \cite{county}. In this regard, it was necessary to 
consider the 38,815 largest entry of the doubly-stochastic matrix 
to complete the hierarchical 
clustering of the 3,140 counties. The probability is 0.973469 that a 
random digraph with 3,140 nodes and 38,414 links is strongly connected
\cite[p. 361]{karonski},
 where
$0.973469=e^{-2 e^{-4.30917}}$,
and $38,814= 3140(\log{3140} +4.30917)$. 
Evidence of systematic structure in the migration
flows can, thus, be adduced, since the digraph based on the 38,814 greatest-valued links was {\it not} strongly connected \cite[p. 8]{county} 
(cf. \cite{killough}). 

In a random digraph with a large number of nodes, the probability is close to 
one that all nodes are either isolated of lie in a single (``giant'') strong
component. The existence of intermediate-sized clusters is thus evidence of
non-randomness, even if such groups are not themselves 
significant according to the 
isolation (difference-of-ranks) criterion of Ling \cite{ling}. 
With randomly-generated data and many taxonomic units, one would expect
the two-stage procedure to yield a dendrogram exhibiting complete
chaining. So, although single-linkage clustering is often criticized
for producing chaining, chains can also be viewed 
simply as indications of inherent randomness in the data. 
In contrast to single-linkage clustering, strong component hierarchical
clustering can merge {\it more} than two clusters (children) into one
(parent) node. This serves to explain why fewer clusters (2,245) were
generated in the intercounty migration study than the 3,139 that 
single-linkage (in the absence of ties) would produce.
\subsection{A cluster-analytic isolation criterion}
Dubes and Jain \cite{DJ} provided ``a semi-tutorial review of the 
state-of-the-art 
in cluster validity, or the verification of results from clustering
algorithms''. Among other evaluative standards, they discussed isolation
criteria, which ``measure the distinctiveness or separation or gaps between a
cluster and its environment''.
Such a statistic was
developed and applied in \cite{qq2} in 
order to extract a small proportion of 5,385 clusters (3,140 of
them single units, 673 pairs, 230 triples, 104 quartets,\ldots) 
for detailed examination based on the 
two-stage analysis of the 1965-1970 United States intercounty migration 
table \cite{county}.

The largest value of the isolation criterion,
for all clusters of fewer than 2940 units, was
attained by a region formed by the eight constituent counties of the state of
Connecticut. (Groups formed by the application of the
two-stage procedure to interareal migration data are, as a strong rule,
composed of contiguous areas \cite{qq,tree}.  This occurs even in the
absence of contiguity constraints, reflecting the distance decay of migration.)
The ll,080th largest doubly-standardized entry, 5,666, corresponding to
movement from New Haven to (New York City suburban) 
Fairfield, unified these eight counties (all
row and column sums had been adjusted to 100,000). Not until the 16,047{\it th}
largest doubly-standardized value, 4,085 (the functional linkage 
from Litchfield, Connecticut  
to Berkshire, Massachusetts),
viewing the clustering procedure as an agglomerative one, was Connecticut
absorbed into a larger region.
The isolation criterion for Connecticut is set equal to
\begin{equation}
25.3175 
= - \log{\Big[\Big( (8 \times 7 + 3132 \times 3131)/(3140 \times 3139) \Big)^{(16047-11080)}\Big]}
\end{equation}
The term in large parentheses is the proportion of cells in the $3,140 
\times 3,140$ table associated
with either movement within Connecticut or within the set of 3,132 
complementary
counties (since intracounty flows are not available, a diagonal
correction is made). This term, raised to the power shown, is the probability
(unadjusted for occupied cells) that {\it none} of $4967 = 16047 - 11080$ 
consecutive
doubly-standardized values would correspond to movement between
Connecticut and its complement. Such a Connecticut-complement linkage
could possibly result in a merger: an {\it unobserved} phenomenon. 
(For further details, including maps, discussion  and extensive 
applications of the isolation criterion developed 
to the U. S. intercounty analysis,
see \cite{qq2}.) This isolation score for the cluster formed by
the four counties of Hawaii--discussed above--was 12.21, 
while the District of Columbia had the highest score, 
23.81, for any single county \cite[Table I]{qq2}.

\section{Complementary Network Flow Procedure}
The creative, productive  network analyst M. E. J. Newman 
has written: ``Edge weights in networks have, with some exceptions
\ldots received relatively little attention in the {\it physics} 
[emphasis added]  literature for 
the excellent reason that in any field one is well advised to look at the
simple cases first (unweighted networks). On the other hand, there are 
many cases where edge weights are known for networks, and to ignore them
is to throw out a lot of data that, in theory at least, could help us to 
understand these systems better'' \cite{newman1}. Of course, the 
numerous (mostly, internal migration) applications of the two-stage 
procedure we have discussed 
above have, in fact, been to such weighted (and directed) networks.

In \cite{newman1}, Newman applied the famous Ford-Fulkerson 
{\it max-flow/min-cut theorem} \cite[Chap. 22]{nijenhuis} to
weighted networks (which he mapped onto unweighted {\it multigraphs}). Earlier, this theorem had been used to study 
Spanish \cite{multiterminal}, Philippine \cite{philippine}, and 
Brazilian, Mexican and Argentinian \cite{brazil} 
internal migration, 
US interindustry flows \cite[pp. 18-28]{tree2} \cite{IO2} \cite[sec. III]{gosling} and the international flow of college students \cite{seps} 
(cf. \cite{cor})--all the corresponding flows now being
left unadjusted, that is {\it not} (doubly- nor singly-) standardized.

In this ``multiterminal'' 
approach, the maximum flow and the dual minimum edge cut-sets, 
between {\it all} ordered pairs of nodes are found. Those cuts 
(often few or even {\it null} in number) which partition the 
$N$ nodes nontrivially--that is, into two sets each of cardinality greater than
1--are noted. The set in each such pair with the fewer nodes is regarded
as a nodal cluster (region, in the geographic context). It has the 
interesting, defining property that fewer people migrate into (from) it, as 
a whole, than into (from) its node. In the Spanish context, the 
(nodal) province of 
Badajoz was found to have a particularly large out-migration sphere of
influence, and 
the (Basque) province of 
Vizcaya (site of Bilbao and Guernica), 
an extensive in-migration field \cite{multiterminal}.
In an analysis of 1967 US interindustry transactions based on 468 
industries, among the industries functioning as  nodes
of  {\it production} complexes with large 
numbers of members were: Advertising; 
Blast Furnaces and Steel Mills; Electronic
Components; and Paperboard Containers and Boxes. Conversely, among those 
serving as nodes of {\it consumption} complexes 
were Petroleum Refining and Meat Animals 
\cite{IO2,gosling}.
\section{Concluding Remarks}
The networks formed by the World Wide Web and the Internet have been the focus
of much recent interest \cite{linked}. Their structures are typically 
represented by 
$N \times N$ {\it adjacency} matrices, the entries of which are simply 0 or 1,
rather than nonnegative numbers, as in internal migration and other
flow tables. One might investigate whether the two-stage
double-standardization and 
hierarchical clustering, and the (complementary) 
multiterminal max-flow/min-cut 
procedures we have sought to bring
to the attention of the active body of 
contemporary network theorists, could yield novel insights into these
 and other important modern structures.

Though quite successful, evidently, in 
simultaneously revealing {\it both} hub-like 
and clustering behavior in 
recorded flows, the indicated implementations of the two-stage 
procedure did not address the 
recently-emerging, theoretically-important 
issues of scale-free networks, power-law descriptions, network evolution 
and vulnerability, and 
small-world properties, among others,
that have been stressed by Barab{\'a}si 
\cite{linked} (and his colleagues and many others in the growing field 
\cite{anthology}). 
(For critiques of these matters, see \cite{doyle,alderson}.) 
One might--using the indicated two-stage 
procedure--compare the hierarchical structure of geographic 
areas using internal migration tables at {\it different} levels of 
geographic aggregation
(counties, states, regions...) (cf. \cite{point}).
To again use the example of France, based on a 
1962-68 $21 \times 21$ interregional
table, R{\'e}gion Parisienne was the most hub-like 
\cite[sec. 4.1]{tree} \cite{france}, while using a finer
$89 \times 89$ 1954-62 interdepartmental table, the dyad 
composed of Seine 
(that is Paris and its immediate suburbs) together with the 
encircling
Seine-et-Oise (administratively eliminated in 1964) 
was most cosmopolitan \cite{winchester} 
\cite[sec. 6.1]{tree}. 
(In \cite{point}, `` two distinct approaches to assessing 
the effect of geographic scale on spatial interactions'' were 
developed.)
\section{Afterword}
It might be of interest to describe the 
immediate motivation for 
this particular communication. 
I had done no further work applying the 
methods described above after 1986, being aware of, but not absorbed in recent
developments in network analysis. 
In May, 2008, Mathematical Reviews asked me to 
review the book of Tom Siegfried \cite{siegfried}, chapter 8 of which is
devoted to the on-going activities in network analysis. This further led 
me (thanks to D. E. Boyce) to the book of Barab{\' a}si \cite{linked}. 
I, then, e-mailed
Barab{\' a}si, pointing out the use of the clustering methodologies 
described above.
In reply, he wrote, in part: ``I guess you were another demo of 
everything being a question of timing-- after a quick look it does 
appear that many things you did have came back as 
questions -- with much more detailed data-- again 
in the network community today. No, I was not aware 
of your papers, unfortunately, and it is hard to know how to get 
them back into the flow of the system.'' 
The present communication 
might be seen as an effort in that direction, alerting
present-day investigators to these demonstratedly fruitful research 
methodologies, and suggesting possible further applications and theoretical 
analysis.
(Additionally, we sent Barab{\' a}si the two-stage analysis 
\cite{romania} for a 1972 
$40 \times 40$ interdistrict migration table for 
his native country, Romania--in 
which the capital of Bucharest was featured as most 
cosmopolitan in nature, and the coupled Black Sea districts of
Constan{\c t}a and Tulcea, as next most. His reply was: 
``Cool-thanks''.)

\begin{acknowledgments}
I would like to express appreciation to the Kavli Institute for Theoretical Physics (KITP)
for technical support.
\end{acknowledgments}

\bibliography{Hub10}

\end{document}